\documentclass[a4paper,aps,prl,showpacs,twocolumn,superscriptaddress]{revtex4-1}   %APS document class style  (PRA,PRL,ETC)
   
\usepackage[utf8]{inputenc}  %Input what you want e.g., é, ł, a, ü
\usepackage[T1]{fontenc}     %Output what you want e.g., é, ł, a, ü
\usepackage[british]{babel}  %Do hyphenation according to british english
\usepackage{lmodern}  %Joe's font
\usepackage[scaled=1.03]{inconsolata} %Monospace font
\usepackage[usenames,dvipsnames]{color} %More color choices
\usepackage[colorlinks,citecolor=blue,linkcolor=magenta,urlcolor=blue]{hyperref}  %Hyperlinks (pink, green, blue)
\usepackage{graphicx} % Package to insert external figures
\usepackage{tikz}
\usepackage[babel]{microtype}  %Improves text justification
\usepackage{amsmath,amssymb,amsthm,bm,mathtools,amsfonts,mathrsfs,bbm,dsfont} %Useful math packages
\usepackage{xspace}  %Useful to add space in macros
\usepackage{multirow}
\usepackage{verbatim}
\usepackage{physics}
\newcommand{\id}{\ensuremath{\mathds{1}}}
\usepackage{bbold}

\newtheoremstyle{mystyle}% name
  {6pt}%Space above
  {6pt}%Space below
  {\normalfont}%Body font
  {0pt}%Indent amount
  {\bf}% Theorem head font
  {.}%Punctuation after theorem head
  { }%Space after theorem head 2
  {}%Theorem head spec (can be left empty, meaning ‘normal’)

\theoremstyle{mystyle}
\newtheorem{theorem}{Theorem}

\newtheorem{observation}{Observation}

\begin{document}
\nonfrenchspacing
\title{Quantifying quantum resources with conic programming}

\author{Roope Uola}
\email{roope.uola@gmail.com}
\affiliation{Naturwissenschaftlich-Technische Fakult\"at, Universit\"at Siegen, Walter-Flex-Str. 3, D-57068 Siegen, Germany}
\affiliation{D\'{e}partement de Physique Appliqu\'{e}e, Universit\'{e}  de Gen\`{e}ve, CH-1211 Gen\`{e}ve, Switzerland}
\author{Tristan Kraft}
\affiliation{Naturwissenschaftlich-Technische Fakult\"at, Universit\"at Siegen, Walter-Flex-Str. 3, D-57068 Siegen, Germany}
\affiliation{SUPA and Department of Physics, University of Strathclyde, G40NG Glasgow,
United Kingdom}
\author{Jiangwei Shang}
\affiliation{Beijing Key Laboratory of Nanophotonics and Ultrafine Optoelectronic Systems, School of Physics, Beijing Institute of Technology, Beijing 100081, China}
\author{Xiao-Dong Yu}
\affiliation{Naturwissenschaftlich-Technische Fakult\"at, Universit\"at Siegen, Walter-Flex-Str. 3, D-57068 Siegen, Germany}
\author{Otfried G\"uhne}
\affiliation{Naturwissenschaftlich-Technische Fakult\"at, Universit\"at Siegen, Walter-Flex-Str. 3, D-57068 Siegen, Germany}

\date{\today}  %Date today

\begin{abstract}
Resource theories can be used to formalize the quantification and manipulation 
of resources in quantum information processing such as entanglement, asymmetry 
and coherence of quantum states, and incompatibility of quantum measurements. 
Given a certain state or measurement, one can ask whether there is a task in 
which it performs better than any resourceless state or measurement. Using 
conic programming, we prove that any general robustness measure (with respect 
to a convex set of free states or measurements) can be seen as a quantifier of 
such outperformance in some discrimination task. We apply the technique 
to various examples, e.g. joint measurability, POVMs simulable by projective 
measurements, and state assemblages preparable with a given Schmidt number.
\end{abstract}

\maketitle

{\it Introduction.---}
In recent years it has become evident that quantum mechanical devices can 
outperform classical ones in tasks like computation, cryptography or 
metrology. Still, it is not entirely clear which quantum mechanical effects are 
responsible for the quantum advantage, and several candidates, such as quantum 
entanglement, Bell nonlocality, quantum contextuality and quantum coherence have 
been discussed~\cite{vidal03, brukner04, curty04, howard14, 
Streltsov_CoherenceRev_2017}. Many phenomena play a role and one cannot expect 
a single phenomenon to be responsible for all applications. So, it is more 
precise to consider a given quantum resource, such as a certain quantum state 
or measurement, and ask: Is there a task in
which this resource outperforms all classical strategies? A general treatment 
of this question leads to the notion of resource theories~\cite{Brandao_RevFramework_QRT_2015, Gour_QRT, Chitambar_QRT_2018, Regula_ConvexGeometry_2018, Liu_2017}, 
where a certain set of states and operations are free and then one can ask for the 
usefulness of the non-free states and operations.

In this paper we present a general method to find tasks in which certain properties of quantum states and measurements provide an advantage. We start by discussing 
the incompatibility of measurements as a resource, and identify a corresponding 
task in terms of a state discrimination problem. Motivated by this, we recognize 
that this result is not limited to quantum incompatibility, as it can be identified 
as a manifestation of a much more general theory, namely, the duality theory of 
conic programming~\cite{Cone_programming_Gaertner}.

Conic programming is a branch of convex optimization that includes 
linear and semi-definite programming (SDP) as special cases. The 
power of introducing this method in our framework lies in the 
fact that, whereas examples such as incompatibility of observables
could be treated with SDPs with specific linear constraints, conic 
programming applies to more general structures. This leads to
task-oriented formulations for various measures of quantumness in 
cases where the linear or semidefinite constraints are harder to 
write down (e.g., coexistence) or even when the constraints are not 
known (e.g., simulability and assemblages related to certain Schmidt number states). 
As a consequence, finding task-oriented characterizations for 
non-classical sets of measurements or assemblages is possible 
in one go.

To demonstrate the general applicability of our approach,
we consider four different scenarios. First, in 
Ref.~\cite{Heinosaari_Postmeasurement_Information_2018} it was 
shown that incompatibility of measurements is necessary in order 
to gain advantage from prior information in state assemblage discrimination 
tasks. We show that also the reverse implication holds, namely that 
for any set of incompatible measurements there exists an instance of 
state assemblage discrimination in which prior information provides an 
advantage. 

Second, we discuss the outperformance of projective or von Neumann measurements 
(PVMs) by generalized measurements or positive operator valued measures (POVMs). More precisely we show that for any POVM that is not 
simulable by PVMs there exists a state discrimination task in which the outperformance 
becomes evident.

Third, it was shown in Ref.~\cite{Piani_Watrous_All_Entangled_States} that all 
entangled states provide an advantage in channel discrimination. This result was 
refined in Ref.~\cite{Piani_moreEnt_betterPerformance_2018} where it was shown 
that higher Schmidt number implies better performance in channel discrimination 
tasks, see also Ref.~\cite{Piani_Watrous_Char_of_EPR_Steering}. Our approach 
implies that preparation of state assemblages from states with higher Schmidt 
number also leads to better performance in tailored subchannel discrimination 
tasks. Such tasks for Schmidt number one, i.e. steering, have been experimentally implemented~\cite{Sun18}. Specifically, this results in semi-device independent Schmidt number 
witnesses.

Fourth, we connect state robustnesses with state ensemble robustnesses. 
This gives observable measures for state resources that can be implemented 
using basic phase estimation protocols as has been done experimentally in 
the case of coherence~\cite{Fei_Experiment_Robustness_2018}. In fact, such 
experimental consequences are generic in our approach: The dual of a conic 
program results in observable witnesses, making an experimental verification of 
the resource character feasible.

{\it Minimum-error state discrimination.---}
A fundamental task in quantum information theory is that of minimum-error state discrimination. Suppose we are given a quantum state $\varrho_a$ from some ensemble $\mathcal{E} = \qty{p_a, \varrho_a}_a$ with prior probabilities $p_a$. Our task is to find a POVM $\mathbf{M}=\qty{M_a}_a$, i.e. a set of positive operators summing up to identity, that gives the best probability of guessing the index $a$ correctly. Here we interpret the outcomes $a$ of the measurement as our guesses. In other words, we are interested in maximizing the quantity $p_{\text{guess}}(\mathcal{E}) = \sum_a p_a\tr[\varrho_a M_a]
$ over all measurements $\qty{M_a}_a$.

A typical instance of this problem is called state discrimination with post-measurement information~\cite{Ballester_Postmeasurement_Information_2008, Gopal_Postmeasurement_Information_2010}. The ensemble $\mathcal{E} = \qty{p_a, \varrho_a}_{a\in I}$ can be partitioned into nonempty disjoint ensembles $\mathcal{E}_x = \qty{p_a, \varrho_a}_{a\in I_x}$, where $\bigcup_x I_x=I$. The label $x$ is revealed after performing the measurement $\mathbf{M}$, as it is the case in the BB84 protocol in quantum key distribution. This additional information cannot decrease the probability of guessing correctly. The success probability can be increased even more by providing this information prior to the measurement, since then one can tailor a separate measurement for each label $x$ individually. Hence, in general it holds that $p_{\text{guess}}(\mathcal{E})\leq p_{\text{guess}}^{\text{post}}(\mathcal{E})\leq p_{\text{guess}}^{\text{prior}}(\mathcal{E})$. It was proven in Ref.~\cite{Heinosaari_Postmeasurement_Information_2018} that $p_{\text{guess}}^{\text{post}}(\mathcal{E})= p_{\text{guess}}^{\text{prior}}(\mathcal{E})$ if and only if there exist compatible measurements that maximize the success probabilities in these tasks.

{\it Incompatibility provides an advantage in state discrimination with prior information.---}
To illustrate our main idea we start by showing that the connection found in Ref.~\cite{Heinosaari_Postmeasurement_Information_2018} can be refined in the sense that for every set of incompatible measurements there exists a state discrimination task in which it performs better than any compatible set.
A measurement assemblage, i.e. a collection of POVMs, $\mathcal{M} = \qty{\mathbf{M}_x}_x = \{M_{a|x}\}_{a,x}$ is called \emph{compatible} or \emph{jointly measurable} (JM), if there exist probability distributions $p(\cdot|x,\lambda)$ and a joint POVM $\mathbf{G}=\qty{G_{\lambda}}_{\lambda}$ such that $M_{a|x}=\sum_{\lambda} p(a|x,\lambda)G_{\lambda}$. Otherwise, the collection is called \emph{non-jointly measurable} or \emph{incompatible}. This definition has a clear operational interpretation. One can collect the statistics of the POVM $\qty{G_{\lambda}}_{\lambda}$ and obtain the statistics of the $\qty{\mathbf{M}_x}_x$ by classical post-processing.

\begin{figure}
    \centering
    \includegraphics[scale=1]{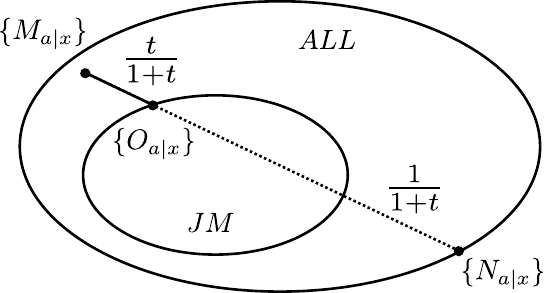}
    \caption{Geometrical interpretation of the incompatibility robustness. Given a set of POVMs $\{M_{a|x}\}$ we search for another set of measurements $\{N_{a|x}\}$ such that the smallest mixture results in a compatible set of measurements $\{O_{a|x}\}$.}
    \label{fig:JMRobustness}
\end{figure}

A natural quantifier of incompatibility is the so-called incompatibility robustness ($\mathcal{IR}$)~\cite{Uola15} 
\begin{equation}\label{Eq:POVMdecomposition}
    \mathcal{IR}(M_{a|x})=\min\qty{t\geq0 \bigg| \frac{M_{a|x}+t N_{a|x}}{1+t}=O_{a|x} \in JM},
\end{equation}
where the optimization is performed over all POVMs $\{N_{a|x}\}_{a,x}$, see also Fig.~\ref{fig:JMRobustness}. The incompatibility robustness can be cast as the following SDP~\cite{Uola15}
\begin{eqnarray}
1+\mathcal{IR}=\min_{\tilde{G}_{\lambda}}\, && \sum_{\lambda}\frac{\tr [\tilde{G}_{\lambda} ]}{d} \\
\text{s. t.: }&& \sum_{\lambda}D(a|x,\lambda) \tilde{G}_{\lambda} \geq M_{a|x} \text{ for all }a,x \notag\\
&& \sum_{\lambda} \tilde{G}_{\lambda} = \frac{\id_d}{d} \sum_{\lambda} \tr[\tilde{G}_{\lambda}], \quad \tilde{G}_{\lambda}\geq 0. \notag
\end{eqnarray}
The dimension of the space is $d$, $D(a|x,\lambda)$ are deterministic post-processings~\cite{Ali09}, $\tilde{G}_{\lambda} = (1+t)G_{\lambda}$ and $G_{\lambda}$  is a joint POVM of $\qty{O_{a|x}}_{a,x}$. As strong duality holds~\cite{Uola15}, the solutions of the primal and dual problems coincide. Computing the dual and picking an optimal choice for the dual variables $\{Y^{a|x}\}_{a,x}$ (see Appendix) one gets
\begin{equation}
\label{Eq:numerator}
    \sum_{a,x} \tr[M_{a|x}Y^{a|x}]=1+\mathcal{IR}.
\end{equation}

Denoting a state ensemble as $\mathcal{E}=\qty{p(x)p(a|x),\varrho_{a|x}}$, where $p(x)$ is the probability of being in the sub-ensemble $x$ and $p(a|x)$ denotes the probability of the label $a$ within the sub-ensemble $x$, allows us to upper bound the success probability for the set $\{M_{a|x}\}_{a,x}$ of POVMs as
\begin{align}
    p_{\text{succ}}(\mathcal{M}, \mathcal{E}) &= \sum_{a,x}p(a,x)\tr[M_{a|x}\varrho_{a|x}] \notag \\
    &\leq (1+\mathcal{IR}) \max_{JM} p_{\text{succ}}(\qty{O_{a|x}},\mathcal{E}).
\end{align}
The maximization is taken over jointly measurable sets of POVMs $\qty{O_{a|x}}_{a,x}$. For the inequality we have used Eq.~\eqref{Eq:POVMdecomposition}. Rewriting this gives
\begin{equation}
\label{Eq:PsuccRatio}
    \frac{p_{\text{succ}}(\mathcal{M}, \mathcal{E})}{\max_{JM} p_{\text{succ}}(\qty{O_{a|x}},\mathcal{E})} \leq 1+\mathcal{IR}.
\end{equation}
The dual variables in Eq.~\eqref{Eq:numerator} are positive semi-definite matrices and one can obtain $Y^{a|x}/\tr[Y]=p(x)p(a|x)\varrho_{a|x}$, where $\tr[Y]=\sum_{a,x}\tr[Y^{a|x}]$. Hence, the left hand side of Eq.~\eqref{Eq:numerator} is, up to a factor, the success probability in a state discrimination task with prior information. Inserting the optimal $Y^{a|x}$ into Eq.~\eqref{Eq:PsuccRatio} and noting that for jointly measurable sets the denominator in Eq.~\eqref{Eq:PsuccRatio} is less than or equal to one we arrive at the following observation.
\begin{observation}\label{Obs:JMisusefull}
For any set of incompatible POVMs $\qty{M_{a|x}}$ there exists a state discrimination task with prior information such that
\begin{equation}
\label{Eq:Observation1}
    \sup_{\mathcal{E}}\dfrac{p_{\text{succ}}(\qty{M_{a|x}}, \mathcal{E})}{\max_{JM} p_{\text{succ}}(\qty{O_{a|x}},\mathcal{E})} = 1+\mathcal{IR}(M_{a|x}).
\end{equation}
\end{observation}
The above Observation can be seen as a semi-device independent statement about measurement incompatibility. Namely, if we can trust the preparation device, i.e. trust $\mathcal{E}$, then we can certify the incompatibility of measurements without assuming anything about their specific form.

In the following sections we show that statements similar to Observation~\ref{Obs:JMisusefull} can be made for any convex and compact subset of POVMs using conic programming.

{\it Conic programming.---}
A subset $C$ of a vector space $V$ is called a \emph{convex cone} if it is convex and for any ${x\in 
C}$ one has ${ax\in C}$ for all $a\geq 0$. The \emph{dual cone} $C^*$ is defined as $C^*=\qty{y \vert 
\braket{x}{y}\geq 0~\forall x\in C}$. Consider a cone program~\cite{Cone_programming_Gaertner}
\begin{eqnarray}
\label{Eq:ConicProgram}
\max_X\, &&\, \tr[AX]  \\
\text{s. t.: }&& \Lambda[X] \leq B,\quad X \in C, \notag
\end{eqnarray}
where $\Lambda$ is a linear operator and $\geq$ denotes the partial order in the positive semi-definite cone of operators. Using Lagrange duality the dual cone program reads
\begin{eqnarray}
\label{Eq:ConicProgramDual}
\min_Y\, && \tr[BY] \\
\text{s. t.: }&& \Lambda^{\dagger}[Y] - A\in C^*, \quad Y \geq 0. \notag
\end{eqnarray}
As in the case of SDPs, strong duality holds if and only if Slater's conditions are fulfilled and the primal problem is finite~\cite{Cone_programming_Gaertner}. In our scenarios Slater's conditions reduce to $B-\Lambda[X]>0$.

{\it Generic robustness measures and state discrimination.---}
Label the set of measurement assemblages with a fixed number of inputs and outputs by $S$. In our discussion a free set $F$ of POVMs is a convex and closed subset of $S$. For a set of POVMs $\{M_{a|x}\}_{a,x}$ in $S$ we can define a generalized robustness with respect to $F$ as
\begin{equation}
\mathcal{R}_F(M_{a|x}) = \min\qty{t\geq0 \bigg| \frac{M_{a|x}+t N_{a|x}}{1+t}=O_{a|x}\in F}.
\end{equation}
A crucial difference to incompatibility robustness is that $F$ is a generic convex and compact subset of $S$, and does not need to be characterizable by an SDP. The generalized robustness can be cast as the following optimization problem
\begin{eqnarray}
\min_t\, && 1+t \\
\text{s. t.: }&&  \frac{M_{a|x}+tN_{a|x}}{1+t}=O_{a|x}\in F, \quad \qty{N_{a|x}}\in S,\quad t\geq 0. \notag
\end{eqnarray}
Defining new variables $\tilde{O}_{a|x}=(1+t) O_{a|x}$ allows writing the above problem as a cone program
\begin{eqnarray}
\min_{\tilde{O}_{a|x}}\,&& \frac{1}{\abs{x}} \sum_{a,x} \frac{\tr[\tilde{O}_{a|x}]}{d} \\
\text{s. t.: }&& \tilde{O}_{a|x} \geq M_{a|x}, \ \tilde{O}_{a|x} \in C_F, \notag
\end{eqnarray}
where $\abs{x}$ is the number of inputs and $C_F$ is the conic hull of $F$. The detailed derivation can be found in the Appendix. The dual program reads
\begin{eqnarray}
\max_{Y^{a|x}}\, && \sum_{a,x} \tr[M_{a|x}Y^{a|x}] \\
\text{s. t.: }&& Y \geq 0, \quad \tr[YT]\leq 1\, \forall\ T\in F, \notag
\end{eqnarray}
where the dual variable $Y=\text{diag}[Y^{a|x}]$ is block-diagonal.

Note that any point in $C_F$ with full rank either fulfils Slater's conditions or it can be scaled up, i.e. multiplied with a sufficiently large positive number, to a point that does. In our examples all cones have a full rank point. Hence, from here on our results have the implicit assumption that the set $F$ is such that the related cone programs fulfil Slater's conditions. 

Similar reasoning as before results in our main 
theorem.
\begin{theorem}\label{Mainresult}
Let $F$ be a convex and compact set of measurement assemblages. For any collection of POVMs $\{M_{a|x}\}_{a,x}$ that is not in $F$ there exists an instance of state discrimination, where $\{M_{a|x}\}_{a,x}$ strictly outperforms all sets of POVMs in $F$. The outperformance is exactly quantified by the generalized robustness with respect to $F$, i.e.,
\begin{equation}
\label{Eq:Theorem1}
    \sup_{\mathcal{E}}\dfrac{p_{\text{succ}}(\qty{M_{a|x}}, \mathcal{E})}{\max_{F} p_{\text{succ}}(\qty{O_{a|x}},\mathcal{E})} = 1+\mathcal{R}_F(M_{a|x}).
\end{equation}
\end{theorem}

Similarly to the case of incompatibility robustness the above result can be seen as a semi-device independent statement about the properties of the measurements. 

As our result applies to any convex and compact set of free POVMs, the question remains to characterize some interesting sets. To give an example, one could go along the lines of joint measurability and take the slightly more general set of coexistent POVMs~\cite{Reeb13}. Note that coexistence is equivalent to the joint measurability of all binarizations and, hence, it can be formulated as an SDP. However, the cone formulation allows one to prove the connection to state discrimination without specifying this SDP. To fulfil Slater's conditions one can take uniform POVMs and scale them up to operators that are larger than identity.

To give an example of a situation where an SDP formulation is not known we consider simulability of POVMs. Recall that state discrmination provides a celebrated example of a task in which POVMs can perform better than PVMs~\cite{Ivanovic87,Peres88}. This statement can be hardened by considering the subset of POVMs that is simulable with all PVMs as the free set $F$ in Theorem~\ref{Mainresult}. This set is defined as those POVMs that can be written in the form $M_{a}=\sum_j p(j) \sum_i p(a|i,j) P_{i|j}$, where $p(\cdot)$ and $p(\cdot|i,j)$ are probability distributions, and $\{P_{i|j}\}$ exhaust the set of projective measurements. In Ref.~\cite{Oszmaniec17} this set was shown to coincide with the convex hull of PVMs and in Ref.~\cite{Oszmaniec18} it was shown that one can reach all measurements by allowing postselection. Note that for a fixed set of simulators, e.g. all PVMs or all binary POVMs, the set is convex. Concerning compactness, one can argue that if a simulable set were not compact, then one could close it as this set can approximate arbitrarily well its own closure. It is worth noting that simulability can also be defined for measurement assemblages~\cite{Guerini17} and that our formalism applies to this scenario given that the free set is convex and compact, which can be achieved by taking the convex hull and the closure if necessary.

{\it Robustness of state assemblages and subchannel discrimination.---}
In Ref.~\cite{Piani_Watrous_Char_of_EPR_Steering} it was shown that for any steerable state assemblage there exists a one-way LOCC assisted subchannel discrimination task in which the assemblage outperforms all unsteerable ones. Here we show that such behaviour is not specific to the case of steering, but it is rather a generic feature of convex and closed sets of assemblages.

In a subchannel discrimination task one aims at discriminating between different elements of an instrument $\mathbf{\Lambda}=\{\Lambda_a\}_a$, i.e. a collection of completely positive maps that sums up to a trace preserving map, with some POVM $\mathbf{N}$. For a given quantum state $\varrho$ the success probability reads  
\begin{align}\label{Eq:subchandiscr}
    p_{\text{succ}}(\varrho,\mathbf{\Lambda},\mathbf{N})=\sum_{a}\text{tr}[\Lambda_a(\varrho)N_a].
\end{align}
For a state assemblage $\{\varrho_{a|x}\}_{a,x}$ we define similarly the success probability as $p_{\text{succ}}(\{\varrho_{a|x}\},\mathbf{\Lambda},\mathbf{N})=\sum_{a,x}\text{tr}[\varrho_{a|x}\Lambda_a^\dagger(N_x)]$. This can be seen as the probability of correctly guessing the subchannel with the assistance of one-way LOCC. Namely, Bob performs a measurement $\mathbf{N}$, communicates the outcome to Alice, she then performs the corresponding measurement and reports the outcome $a$ as the guess. Note that we assume Alice's measurements and the shared state to be such that they prepare the assemblage $\{\varrho_{a|x}\}_{a,x}$.

One can define generalized robustnesses for state assemblages and formulate them through conic programming as in the case of measurement assamblages. The only difference is the normalization and, hence, the interpretation of the dual program. For measurement assemblages the dual can be identified as a state disrcimination problem and for state assemblages the dual corresponds to a subchannel discrimination task (see Appendix). We arrive at the following Theorem.

\begin{theorem}\label{Mainresult2}
Let $F$ be a convex and compact set of state assemblages. For any state assemblage $\{\varrho_{a|x}\}_{a,x}$ that is not in $F$ there exists an instance of (one-way LOCC assisted) subchannel discrimination, where $\{\varrho_{a|x}\}_{a,x}$ strictly outperforms all assemblages in $F$. The outperformance is exactly quantified by the generalized robustness with respect to $F$, i.e.,
\begin{align}
    \sup_{\mathbf{\Lambda},\mathbf{N}}\dfrac{p_{\text{succ}}(\{\varrho_{a|x}\},\mathbf{\Lambda},\mathbf{N})}{\max_{F} p_{\text{succ}}(\qty{\sigma_{a|x}},\mathbf{\Lambda},\mathbf{N})} = 1+\mathcal{R}_F(\varrho_{a|x}).
\end{align}
\end{theorem}

To give a physically motivated example of the free set $F$, we consider 
assemblages that can be prepared using states with Schmidt number $n$ or 
smaller. As in the case of measurement simulability, an SDP 
formulation for such scenario is not known. For the proof of convexity and compactness of these free sets, we refer to the Appendix. Slater's conditions are fulfilled as the cones include a full rank point (e.g.  the uniform 
assemblage). It is worth mentioning that the inclusion of the free set for the 
case of Schmidt number $n$ is proper to that of Schmidt number $n+1$~\cite{Hulpke04}. This example is in the spirit of Ref.~\cite{Piani_moreEnt_betterPerformance_2018}, where it was shown that higher Schmidt number provides an advantage in channel discrimination tasks. Moreover, the example refines the characterization of steerable assemblages given in Ref.~\cite{Piani_Watrous_Char_of_EPR_Steering}, hence, leading to a semi-device independent approach to Schmidt number verification.

{\it State ensembles.---}
The connection between robustness and discrimination in the case of state ensembles follows from the discussion on state assemblages by setting $x=1$. This corresponds to the case in Eq.~(\ref{Eq:subchandiscr}). To give a physically motivated example, we consider ensembles that are created through a given instrument $\mathbf{\Lambda}$.

To make a connection to the robustness of a given state denoted as $\mathcal{R}(\varrho)$ (may it be, e.g., entanglement, coherence, asymmetry or coherence number robustness), we note that when operating only within the set $\{\{\Lambda_a(\varrho)\}_a|\varrho\in\mathcal S(\mathcal H)\}$ of ensembles [here $S(\mathcal H)$ denotes the set of positive unit trace operators], we can define the robustness of an ensemble $\{\varrho_a\}_a$ as $\mathcal{R}(\varrho_a) = \min\qty{t\geq 0 | \varrho_a+t\tau_a=(1+t)\sigma_a\in C_F}$, where $F$ is the set of ensembles preparable with the given instrument and resourceless states, and $\{\tau_a\}_a$ is any ensemble preparable with the given instrument. The techniques presented in the previous section give a subchannel discrimination problem as the dual of the robustness, with the dual variables being POVMs.

To fulfil Slater's conditions we need a full rank point in $F$. Typical free sets include the maximally mixed state or the maximally mixed ensemble and, hence, the set $F$ has a full rank point.

The ensemble robustness is always less than or equal to the state robustness as one can input an optimal solution of the state robustness to the instrument. We have $p_{\text{succ}}(N_a, \varrho_a)  \leq (1+\mathcal{R(\varrho)}) \max_{F} p_{\text{succ}}(N_a,\sigma_a)$, where $\{N_a\}_a$ is a POVM. Whenever the instrument is a bijection from the set of states to the set of ensembles, e.g. in phase discrimination, the ensemble robustness coincides with the corresponding state robustness. Therefore, maximizing over all instruments and POVMs saturates the bound (see also Theorem~\ref{Mainresult2}).

We have recovered the result of Ref.~\cite{Adesso_Operational_Advantage_Subchannel} stating that robustnesses of state resources are connected to subchannel discrimination. In contrast to the former result in which a witness was split into an instrument and a POVM, our construction can use, for example, any phase estimation protocol as the instrument and the witness is simply a POVM. It is worth mentioning that phase discrimination has been used to measure the robustness of coherence~\cite{Fei_Experiment_Robustness_2018} in a recent experiment.

{\it Conclusions.---}
In this work we have shown how various optimal and non-optimal witnesses for the classical to quantum border can be written in an observable form using conic programming. These witnesses arise from generalized robustnesses and as such the results open up the possibility to define observable quantifiers for the quantum properties that are within the reach of current experiments~\cite{Fei_Experiment_Robustness_2018, Sun18}.

In comparison to earlier efforts in this direction, our techniques apply to any properties of measurement and state assemblages that form a convex and compact subset, whereas 
former techniques have dealt with single properties such as quantum steering~\cite{Piani_Watrous_Char_of_EPR_Steering} or 
with sets of properties related to single states~\cite{Adesso_Operational_Advantage_Subchannel}. This allowed us not only to answer open 
questions~\cite{Piani_moreEnt_betterPerformance_2018} and to push forward earlier works~\cite{Heinosaari_Postmeasurement_Information_2018}, but also to develop novel methods 
in the field of semi-device independent quantum information processing and to attack the question whether POVMs provide an advantage over PVMs in a quantitative way.

For future research it will be interesting to identify other properties than the ones discussed here as the free set. Also, the question of generalizing the results to the level of quantum channels and instrument assemblages might provide new insights to the properties of these notions, e.g. in the resource theory of quantum memories~\cite{Rosset_Resource_Theory_Q_Memories_2018}. Finally, the operations that do not generate resources from the free set cannot increase the robustness measure. Thus, it would be interesting to characterize the physical interpretation of these
operations and the properties of the robustness measure under time evolutions or classical pre- and post-processing.

\begin{acknowledgments}

We would like to thank Marco Piani and Leonardo Guerini for discussions. This work was 
supported by the DFG and the ERC (Consolidator Grant 683107/TempoQ).
J.S. acknowledges support by the Beijing Institute of 
Technology Research Fund Program for Young Scholars and the National Natural 
Science Foundation of China through Grant No. 11805010. X.D.Y. acknowledges 
funding from a CSC-DAAD scholarship. R.U. is thankful for the support from the Finnish Cultural Foundation.

{\it Note added.---} During the preparation of the manuscript we became aware of some related works. In Ref.~\cite{Carmeli18} Carmeli et al. show, using a different method, that incompatibility can always be detected by a state discrimination task with partial intermediate information. In particular they prove that any linear incompatibility witness can be implemented by some state discrimination task. In Ref.~\cite{Skrzypczyk19} Skrzypczyk et al. also prove the quantitative connection between the incompatibility robustness and the outperformance of compatible measurements by incompatible ones in tailored state discrimination tasks. Moreover, they show the completeness of state discrimination games as resource monotones, thus completely characterizing the partial order in a resource theoretical sense. Furthermore we became aware of two other related works, one by Oszmaniec and Biswas~\cite{Oszmaniec19}, and another one by Takagi and Regula~\cite{Regula19}.

\end{acknowledgments}

\appendix

\section{Appendix A: Incompatibility robustness SDP and its dual}

The incompatibility robustness $1+\mathcal{IR}$, can be cast as the following SDP
\begin{eqnarray}
\min_{\tilde{G}_{\lambda}}\,&& \sum_{\lambda}\frac{\tr [\tilde{G}_{\lambda} ]}{d} \\
\label{Eq:constraint1}\text{s. t.: }&& \sum_{\lambda}D(a|x,\lambda) \tilde{G}_{\lambda} \geq M_{a|x} \text{ for all }a,x \\
\label{Eq:constraint2} && \tilde{G}_{\lambda}\geq 0 \\
&& \sum_{\lambda} \tilde{G}_{\lambda} = \frac{\id_d}{d} \sum_{\lambda} \tr[\tilde{G}_{\lambda}]\notag,
\end{eqnarray}
where $d$ is the dimension of the space of the $G_{\lambda}$ such that 
$\sum_{\lambda} G_{\lambda} = \id_d$. The number of labels $a$ we denote by $|a|$, and similarly $|x|$ denotes the number of labels $x$. The number of constraints in Eq.~\eqref{Eq:constraint1} is $|a|\cdot |x|$ and in Eq.~\eqref{Eq:constraint2} it is $|a|^{|x|}$.
This SDP has both, equalities and inequalities as constraints and it is of the general form
\begin{eqnarray}
\min_{X}&& \langle A,X \rangle \\
\text{s. t.: }&& \Phi(X)=B_1 \notag \\
&& \Psi(X)\geq B_2 \notag \\
&& X\geq 0. \notag
\end{eqnarray}
We choose $A=\id_{d\cdot |a|^{|x|}} \qquad\text{and}\qquad 
X=\text{diag}\qty[\tilde{G}_{\lambda}/d]_{\lambda}\in\mathcal{M}(\mathbb{C}^{d\cdot 
|a|^{|x|}})$, which is a block diagnoal matrix with the submatrices 
$\tilde{G}_{\lambda}/d$. The objective function then reads $\langle A,X \rangle 
= \tr[X]=\sum_\lambda \tr [X_\lambda]$.  For the equality constraint we choose 
$B_1=0$ and define a mapping $\Phi: \mathcal{M}(\mathbb{C}^{d\cdot |a|^{|x|}})\mapsto 
\mathcal{M}(\mathbb{C}^{d})$ by
\begin{equation}
    \Phi(X) = \id_d \tr[X]-d \sum_\lambda X_\lambda.
\end{equation}
For the inequality constraint we choose $B_2=\text{diag}\qty[M_{a|x}]_{a,x}$ and we define $\Psi(X):  \mathcal{M}(\mathbb{C}^{d\cdot |a|^{|x|}})\mapsto \mathcal{M}(\mathbb{C}^{d\cdot |a|\cdot |x|})$ by
\begin{equation}
    \Psi(X) = \text{diag}\qty[d\sum_{\lambda}D(a|x,\lambda) X_{\lambda}]_{a,x}.
\end{equation}
The dual problem then reads~\cite{JWatrous_Notes_2011}
\begin{eqnarray}
\max_{Z,Y}&& \langle B_1,Z \rangle + \langle B_2,Y \rangle \\
\text{s. t.: }&& \Phi^{\dagger}(Z) + \Psi^{\dagger}(Y) \leq A \notag\\
&& Z\text{ is Hermitian} \notag \\
&& Y\geq 0. \notag
\end{eqnarray}
The dual $\Psi^{\dagger}(Y)$ is straight forward~\cite{Piani_Watrous_Char_of_EPR_Steering}, that is
\begin{eqnarray}
    \Psi^{\dagger}(Y) = \text{diag}\qty[d\sum_{a,x}D(a|x,\lambda) Y^{a|x}]_{\lambda}.
\end{eqnarray}
To find $\Phi^{\dagger}(Z)$ we write
\begin{eqnarray}
    \tr[\Phi(X)Z]&=&\tr[\tr(X) Z]-\tr[d\sum_{\lambda} X_{\lambda} Z] \\
    &=& \tr[X\qty{\tr(Z)\id_{d\cdot a^x}-d(Z\oplus Z\oplus \cdots \oplus Z)}] \notag \\
    &=& \tr[X \Phi^{\dagger}(Z)]\notag.
\end{eqnarray}
From this we directly obtain the dual of the robustness SDP as
\begin{eqnarray}
\label{Eq:IRDual}
\max_{Y_{a|x}}&& \sum_{a,x} \tr[M_{a|x}Y^{a|x}]  \\
\text{s. t.: }&& \text{diag}\qty[d\sum_{a,x}D(a|x,\lambda) 
Y^{a|x}]_{\lambda}\notag\\
&&+ \tr(Z)\id_{d\cdot |a|^{|x|}}-d(Z\oplus Z \oplus \cdots \oplus Z) \leq \id_{d\cdot |a|^{|x|}} \notag\\
&& Z \text{ is Hermitian} \notag\\
&& Y\geq 0. \notag
\end{eqnarray}

\section{Appendix B: Upper bound on the success probability for sets on compatible POVMs}

Next, we show that whenever a set of jointly measurable POVMs is used to discriminate the optimal state assemblage $\qty{Y^{a|x}}$, we find that
\begin{eqnarray}
    \sum_{a,x} \tr[O_{a|x}Y^{a|x}] &=& \sum_{a,x,\lambda} D(a|x,\lambda) \tr[J_{\lambda}Y^{a|x}] \\
    &=:& \sum_{\lambda} \tr[J_{\lambda}\tilde{Y}^{\lambda}].
\end{eqnarray}
From the first constraint of the dual program in Eq.~\eqref{Eq:IRDual} we obtain (from each block labeled by $\lambda$) that $\tilde{Y}^{\lambda}\leq\frac{\id_d}{d}(1-\tr Z)+Z$. This leads to
\begin{eqnarray}
    \sum_{\lambda} \tr[J_{\lambda}\tilde{Y}^{\lambda}] &\leq& \sum_{\lambda} \tr[J_{\lambda}\frac{\id_d}{d}(1-\tr Z)+Z] \notag\\
    &=& \tr[\frac{\id_d}{d}(1-\tr Z)+Z] = 1.
\end{eqnarray}
Hence, for any set of jointly measurable POVMs it holds that
\begin{equation}
\label{Eq:PsuccJM}
    \sum_{a,x} \tr[O_{a|x}Y^{a|x}] \leq 1.
\end{equation}

\section{Appendix C: Construction of the state discrimination task with prior information from the optimal dual variable}

From the optimal solution of the dual we also construct a state discrimination task with prior information in the following way. First observe that
\begin{eqnarray}
Y^{a|x} &=& \tr[Y]\frac{\sum_{a'}\text{tr}[Y^{a'|x}]}{\tr[Y]}\frac{\tr[Y^{a|x}]}{\sum_{a'}\tr[Y^{a'|x}]}\frac{Y^{a|x}}{\tr[Y^{a|x}]}\notag\\
&=& \tr[Y]p(x)p(a|x)\varrho_{a|x}.
\end{eqnarray}
Inserting this into the objective function of the dual in Eq.~\eqref{Eq:IRDual} yields
\begin{eqnarray}
    &&\sum_{a,x} \tr[M_{a|x}Y^{a|x}] \notag \\
    &=& \tr[Y] \sum_{a,x} p(x)p(a|x) \tr[M_{a|x}\varrho_{a|x}] \notag \\
    &=& \tr[Y]\quad p_{\text{succ}}(M_{a|x}, \varrho_{a|x}).
\end{eqnarray}
Then, using Eq.~\eqref{Eq:PsuccJM} the ratio of success probabilities in Eq.~\eqref{Eq:PsuccRatio} is lower bounded by
\begin{eqnarray}
    &&\frac{p_{\text{succ}}(M_{a|x}, \varrho_{a|x})}{\max_{O_{a|x}\in JM} p_{\text{succ}}(O_{a|x},\varrho_{a|x})} \\
    &=& \frac{\sum_{a,x} \tr[M_{a|x}Y^{a|x}]}{\max_{O_{a|x}\in JM} \sum_{a,x} \tr[O_{a|x}Y^{a|x}]}\\
    &\geq& \sum_{a,x} \tr[M_{a|x}Y^{a|x}] =1+\mathcal{IR}.
\end{eqnarray}
The inequality follows from Eq.~\eqref{Eq:PsuccJM}. From this, Observation~\ref{Obs:JMisusefull} follows.

%%%%%%%%%%%%%%%%%%%%%%%
%APPENDIX B
%%%%%%%%%%%%%%%%%%%%%%%

\section{Appendix D: Robustness of sets of measurements and conic programming}  

Denote any set of free POVMs by $F$. Let $|x|$ be the number of POVMs. The generalized robustness is defined by
\begin{equation}
    \mathcal{R}_F(M_{a|x}) = \min\qty{t\geq 0 | \frac{M_{a|x}+tN_{a|x}}{1+t}=O_{a|x}\in F}.
\end{equation}
This can be cast as the following conic program
\begin{eqnarray}
1+\mathcal{R}_F(M_{a|x})  = \min_t &&\, 1+t \\
\text{s. t.: }&& t\geq 0 \\
&& \frac{M_{a|x}+tN_{a|x}}{1+t}=O_{a|x}\in F \\
&& \qty{N_{a|x}}\text{ is a POVM}.
\end{eqnarray}
Solving for $N_{a|x}$ one obtains
\begin{eqnarray}
\min_t &&\,1+t \\
\text{s. t.: }&& t\geq 0 \\
&& (1+t) O_{a|x} - M_{a|x} \geq 0 \\
&& O_{a|x} \in F.
\end{eqnarray}
By defining $\tilde{O}_{a|x}=(1+t) O_{a|x}$ this can be written as
\begin{eqnarray}
\min_{\tilde{O}_{a|x}}&& \frac{1}{\abs{x}} \sum_{a,x} \frac{\tr[\tilde{O}_{a|x}]}{d} \\
\text{s. t.: }&& \tilde{O}_{a|x} \geq M_{a|x} \\
&& \tilde{O}_{a|x} \in C_F,
\end{eqnarray}
where $C_F$ is a cone with basis $F$. This can be brought into the form of Eq.~(\ref{Eq:ConicProgram}) by choosing $A=-\frac{1}{\abs{x} d}\id$, $X=\text{diag}(\tilde{O}_{a|x})_{a,x}$, $B=-\text{diag}(M_{a|x})_{a,x}$, $\Lambda=-\textit{id}$. Then, the dual cone program reads
\begin{eqnarray}
\max_{Y^{a|x}}&& \sum_{a,x} \tr[M_{a|x}Y^{a|x}] \\
\text{s. t.: }&& -Y+\frac{1}{|x| d}\id \in C_F^* \\
&& Y \geq 0.
\end{eqnarray}
The first constraint translates to $\braket{\frac{1}{|x| d}\id-Y}{T}\geq 0$. Hence, $\tr[YT]\leq \tr[T/|x| d]$ for all $T\in C_F$ or equivalently $\tr[YT]\leq 1$ for all $T\in F$. The final form of the dual then reads
\begin{eqnarray}
\max_{Y^{a|x}}\,&& \sum_{a,x} \tr[M_{a|x}Y^{a|x}] \\
\text{s. t.: }&& Y \geq 0 \\
&& \tr[YT]\leq 1 \text{ for all } T\in F.
\end{eqnarray}
Note that the last constraint is a typical property of a witness. Similar results hold for state assemblages $\qty{\varrho_{a|x}}_{a|x}$ by simply dropping the factor $1/d$, since the operators $\sum_a\varrho_{a|x}$ have a unit trace.

%%%%%%%%%%%%%%%%%%%%%%%
%APPENDIX C
%%%%%%%%%%%%%%%%%%%%%%%

\section{Appendix E: Robustness of state assemblages and conic programming}

For a given free set of assemblages $F$, an assemblage $\{\varrho_{a|x}\}_{a,x}$, and an subchannel discrimination task $(\mathbf{\Lambda},\mathbf{N})$ we have
\begin{align}
    \dfrac{p_{\text{succ}}(\{\varrho_{a|x}\},\mathbf{\Lambda},\mathbf{N})}{\max_{F} p_{\text{succ}}(\{\sigma_{a|x}\},\mathbf{\Lambda},\mathbf{N})} \leq 1+\mathcal{R}_F(\varrho_{a|x}).
\end{align}
To formulate the statement of Theorem~\ref{Mainresult} for state assemblages we note that the primal problem for the robustness of an assemblage $\{\varrho_{a|x}\}_{a,x}$ with respect to a free set $F$ of assemblages is given as
\begin{eqnarray}
\min_{\tilde{\sigma}_{a|x}}\, && \frac{1}{\abs{x}} \sum_{a,x}\tr[\tilde{\sigma}_{a|x}] \\
\text{s. t.: }&& \tilde{\sigma}_{a|x} \geq \varrho_{a|x}, \quad \tilde{\sigma}_{a|x} \in C_F, \notag
\end{eqnarray}
where $\tilde{\sigma}_{a|x}=(1+t)\sigma_{a|x}$. The dual program can be written as
\begin{eqnarray}
\max_{Y^{a|x}}\,&& \sum_{a,x} \tr[\varrho_{a|x}Y^{a|x}] \\
\text{s. t.: }&& Y \geq 0, \quad \tr[TY]\leq 1\, \forall T\in F. \notag
\end{eqnarray}
We have again denoted by $Y$ the direct sum of the operators $\{Y^{a|x}\}_{a,x}$. Note that Slater's conditions can be verified similarly to the case of measurements for the free sets we are interested in.

Using the techniques introduced in Ref.~\cite{Piani_Watrous_Char_of_EPR_Steering} it is clear that any witness $Y$ of the above form can be cast as a subchannel discrimination task with one-way LOCC measurements. Namely, define subchannels and a POVM as $\Lambda_a^\dagger(|x\rangle\langle x|)=\alpha Y^{a|x}$ and $N_x=|x\rangle\langle x|$, where $\alpha=\|\sum_{a,x} Y^{a|x}\|_\infty^{-1}$ and $\qty{\ket{x}}_x$ is an 
orthonormal basis. If these subchannels do not form an instrument, i.e. $\sum_a\Lambda_a^\dagger(\openone)\neq\openone$, the set can be 
completed into one by defining an extra subchannel as 
$\Lambda(\varrho)=\tr[(\openone-\sum_a\Lambda_a^\dagger(\openone))\varrho]\sigma$, 
where $\sigma$ is some quantum state. It is worth noting that we have one more 
subchannel in the discrimination problem than we have outputs. 

%%%%%%%%%%%%%%%%%%%%%%%
%APPENDIX D
%%%%%%%%%%%%%%%%%%%%%%%

\section{Appendix F: Convexity and compactness of the set of assemblages that can be prepared from states with a fixed Schmidt number}
Convex combinations of such assemblages can be prepared by increasing the size of 
Alice's system. To be more precise, given that Alice's dimension is $d$ and 
that one assemblage is prepared with measurements $\{A_{a|x}\}_{a,x}$ on the 
state $\varrho_{AB}$ and another assemblage with measurements $\{\tilde 
A_{a|x}\}_{a,x}$ on the state $\tilde\varrho_{AB}$, we can consider the convex 
combination 
$\varrho:=\lambda\ketbra{0}{0}\otimes\varrho_{AB}+(1-\lambda)\ketbra{1}{1}\otimes 
\tilde\varrho_{AB}$ and the measurements $\hat A_{a|x}:=\ketbra{0}{0}\otimes 
A_{a|x}+\ketbra{1}{1}\otimes \tilde A_{a|x}$, where $\{\ket{0},\ket{1}\}$ is 
the basis of an auxiliary qubit of Alice.

To prove the compactness of the desired set of assemblages, we first notice that the extremal points are obtained by pure states and that every extremal point can be reached with a finite dimensional Alice. As Bob is assumed to be finite dimensional, any assemblage that is preparable by a Schmidt number $n$ (or smaller) state can be expressed as a finite convex combination of extremal assemblages. Hence, any sequence of assemblages preparable with a Schmidt number $n$ (or smaller) state can be written as
\begin{equation}
(\varrho_{a|x}^m)_m=(\sum_{i=1}^k p_{i|m}\xi_{a|x}^{i|m})_m,
\end{equation}
where $k$ is some fixed finite number depending on the number of dimension 
(Carathéodory's theorem), $p_{i|m}$ is a probability distribution for every 
$m$, and $\xi_{a|x}^{i|m}$ are assemblages preparable with a pure Schmidt rank 
$n$ or less state. The set of assemblages preparable with pure states is 
clearly compact (as it is the image of a cartesian product of compact sets in 
a continuous mapping) and, therefore, for every $i$ we can pick a converging 
subsequence of $\xi_{a|x}^{i|m}$.  Picking the subsequences for different 
indices $i$ sequentially (i.e.  subsequences of subsequences) results in 
a subsequence in which convergence is guaranteed for every $i$. Repeating the 
procedure once more to pick a converging sequence of probability distributions 
gives a subsequence of $(\varrho_{a|x}^m)_m$ that converges. Hence, the set of 
Schmidt number $n$ (or smaller) preparable assemblages is compact.

\bibliography{References}

\end{document}